# Combining experimental and computational methods to unravel the dynamical structure of photoelectrosynthetic interfaces

Matthias M. May[1,2], Wolfram Jaegermann[3]

[1]Universität Tübingen, Institute of Physical and Theoretical Chemistry; [2]Universität Ulm, Institute of Theoretical Chemistry; [3]Technische Universität Darmstadt, Department of Surface Science

matthias.may@uni-tuebingen.de; jaegermann@surface.tu-darmstadt.de

**Abstract**

At photoelectrosynthetic interfaces, an electrochemical reaction is driven by excited charge-carriers from a semiconducting photoabsorber. Structure and composition of this interface determine both the electronic and electrochemical performance of devices, yet this structure is often highly dynamic both in the time-domain and upon applied potentials. We discuss the arising challenges from this dynamical nature and review recent approaches to gain an atomistic understanding of the involved processes, which increasingly involves a combination of experimental and computational methods. Bearing a similarity to solid-electrolyte interphase formation in batteries, their apprehension could help to develop functional passivation layers for high-performance photoelectrosynthetic devices.

## I. Introduction

The semiconductor/electrolyte contact is central to any photoelectrosynthetic device, where the semiconducting absorber is an integral part of the solid/electrolyte interface. With this definition, we include in our discussion buried junctions, but exclude solar-driven electrolysis[1]. This contact may be realized in different chemical and structural complexity including intimate contacts, the involvement of adsorbate/surface states, and chemically formed or deliberately deposited reaction layers. Ideally, chemical or electronic passivation forms active sites involved in charge-transfer for the hydrogen and oxygen evolution reaction (HER/OER), which may also involve additional co-catalysts as films or nanostructures. Furthermore, it must be noted that the interface arrangement and related properties will be changing and must be discriminated after contact formation, reaching electronic equilibrium (in the dark) and for *operando* conditions close to the maximum performance point under illumination (comparable to the maximum power point of solar cells). One should notice that the operation point of running any photoelectrosynthetic device is defined by the needed operational photovoltage times the related operational photocurrent (not by the photocurrent at the reversible redox potential)[2].

Achieved efficiencies of such chemical converter (electrosynthetic, e.g. hydrogen-evolving) devices will be dominated by the photoabsorber component and its performance values (for most reactions, at least a tandem photoabsorber configuration is needed[3,4,2]), by the reactivity of the the outermost reaction layers (in their electrocatalytic performance given by the overvoltages at the operative photocurrents[5]), but also by the contact properties in charge-transfer, charge-trapping, and charge-induced reactions at the different involved junctions. Whereas very many reviews have been published on different aspects of such photoelectrosynthetic devices (too many to cite them here), a detailed, in-depth understanding, specifically on the possibly involved interface effects has not been achieved, yet, and needs further evaluation. Which specific considerations must already be considered in any knowledge-based design of efficient and competitive systems in comparing metal and semiconductor electrolyte junctions was recently discussed in an excellent discussion paper by L. Peter[6]. However, ideal behaviour of the semiconductor was assumed, without involvement of interfacial states, which is probably already for most static systems an oversimplified view. Furthermore, the structure of photoelectrosynthetic interfaces is highly dynamic and changes upon applied potentials, illumination, during the reaction, but also in the course of corrosion.

In the following, we want to address the solid/electrolyte interface of photoelectrosynthetic devices in combining physical and electrochemical viewpoints in the discussion of electrolyte contacts. We will focus on their dynamical (electronic) structure on the timescale of electrochemical reactions and, for select aspects and methods, very briefly outline the current state of the endeavour in developing experimental and computational approaches to gain an atomistic understanding.

## II. Photoelectrosynthetic interfaces

Idealised, high-performance photoelectrochemical regenerative cells are usually taken as basis for the discussion of semiconductor/electrolyte contacts and are often also used when discussing multi-electron transfer reactions with complex electrode compositions. Yet even in the case of ideal semiconductor surfaces, where initially no active surface states within the bandgap exist, dipolar layers can lead to modifications of the work function for different surface orientations (see. Fig. 1a ). These will be further modified by forming a contact to an electrolyte by electrochemical double layers (EDL), depending on the electrolyte composition[7,8,9]. Thus the Fermi level position depends on the orientation and homogeneity and structural arrangement of the solid[11]. If the contact to the electrolyte is established even for this idealized and seldomly given case, the potential drop in the space-charge layer and thus the diffusion voltage for charge-carrier separation as given by the difference of $E_F$ vs. $eU_{red/ox}$ (after contact with the electrolyte) will not be identical to the difference $\Delta\Phi$ of semiconductor vs. electrolyte work functions. These additional dipolar potential drops must be considered as they are likely to change from solid/vacuum to solid/electrolyte interfaces, depending on surface compositions and related surface potentials (as indicated in Fig. 1b[10]).

Furthermore, most semiconductors form active surface states within their bandgaps after loosing bulk translational symmetry in bonding interactions, which must be considered additionally in contact formation. This is shown in Fig. 2, where Shockley surface states close to midgap, as e.g. to be expected for Si-like semiconductors, have been added. For most semiconductors such *intrinsic surface states* are formed already in contact to vacuum and their energetic positions and density will depend on the type of semiconductor, the possibly formed surface reconstructions, and related surface bonds, depending on synthesis and pretreatment[11]. After coming in contact with ambient conditions, added gas phases, and especially to electrolyte solutions, the intrinsic surface states form adsorbate bonds and modified surface compositions and structures, and even reacted surface phases (of nm dimension), which result in *extrinsic surface states* of surface defect states with a modified interfacial density of states. For most semiconductor/electrolyte (sc/el) contacts, these *extrinsic interface states* do exist in varying concentrations and energetic distributions, depending on the history of the semiconductor surface under consideration. Very often the experimentally observed variation of the expected ideal behaviour of semiconductor junctions are related to such "surface states" without any detailed knowledge of their origin and physical properties. These states do not only exist in classical semiconductors, but virtually in all semiconductor materials, including complex metal oxides[12,13,14].

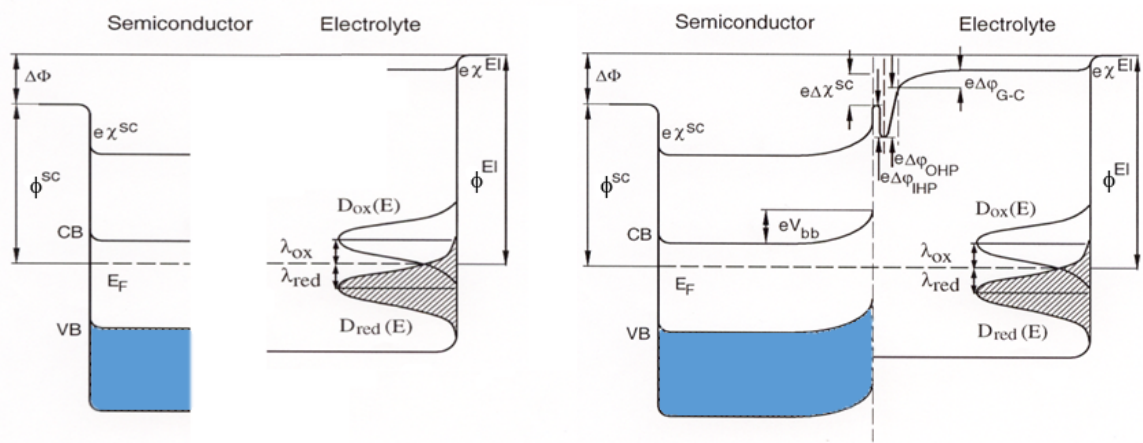

*Figure 1 Contact formation of the semiconductor (SC) with its valence and conduction bands (VB, CB) to the electrolyte (El) leading to an equilibrium Fermi level ($E_F$). The latter contains a well-defined one-electron transfer reversible outer sphere redox couple with its redox potential E(red/ox) given by its work function $\Phi^{el}$, the density of states ($D_{ox}/D_{red}$), and the reorganization energy, λ. Before contact (a), both work functions contain an surface dipole potential drop of eχ, which will be modified after contact (b) formation as schematically shown as contributions of different dipolar or electrochemical double layers (semiconductor dipole, inner Helmholtz, outer Helmholtz (IHP/OHP), Gouy-Chapman (G-C)). As a consequence, the originally given difference in work function after contact formation is divided into an extended space charge layer with a band bending, $eV_{bb}$, and double layer potential drops.*

The situation is even more complex if photoelectrochemical contacts to be used for fuel formation will come into play (see Fig. 2c). The involved multi-electron transfer reactions with high-energy free intermediates need a strong chemical stabilisation of the intermediate species with bond formation to the solid electrode (in the case of HER and OER two and four electrons must be transferred, avoiding the formation e.g. of ·OH and ·H radicals). As a consequence, the bonding of the intermediates under operational conditions will lead to new interface states or even to reacted surface phases with modified energy states[15] if the chemical stabilisation will occur on the semiconductor directly. As a consequence, such reactive surface phases will at first lead to losses of photovoltage due to reduction of the splitting of the quasi-Fermi levels ($_{n/p}E_F$) within the photoabsorber. Additionally, a loss of photocurrent will occur due to increased surface recombination rates. The electrons in the semiconductor will have the chance to recombine efficiently with the holes involving such intermediately formed interface states, as efficient systems must operate close to the maximum power point, strongly reducing space-charge layers. Finally, trapping of charge in these interfaces states may lead to additional overvoltage due to band-edge unpinning as usually observed for non-optimized absorbers. In Fig. 2, we have assumed a close-to-midgap formation of interface states due to the stabilisation of the reaction intermediates of OER, equivalent to charge-trapping of holes for non-optimized electrocatalytic properties. These rearrangement and transfer of reacting holes and electrons at operational conditions define the interface-dependent performance of such devices. Furthermore, 'photoelectrochemical triple-points', where semiconducting absorber/passivation layer, semiconducting or metallic catalyst, and finally the electrolyte meet, can qualitatively change energetic alignment at the interface and hence introduce an additional level of complexity, even under static conditions[16].

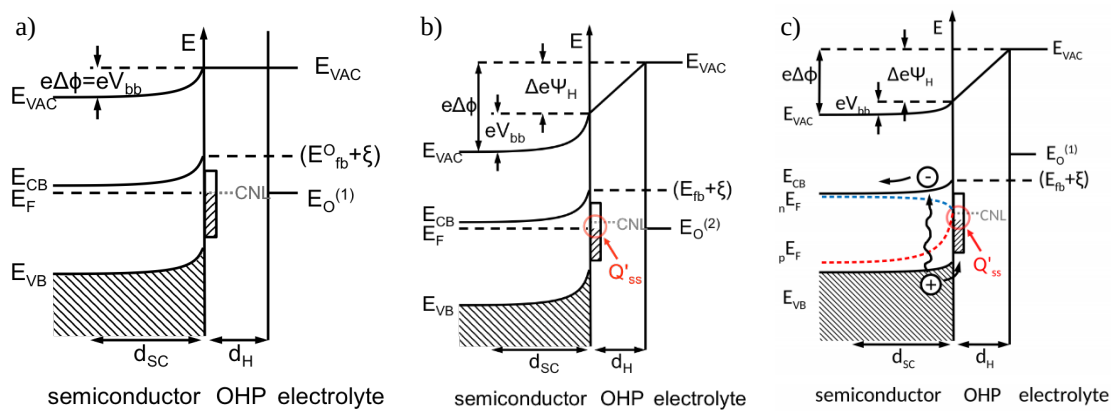

*Figure 2: Influence of surface states on the electric potential drop at semiconductor-electrolyte interfaces (Fermi-level pinning) at two different electrochemical potentials ($E_O^{(1/2)}$). The width of band-bending in the semiconductor and Helmholtz-layer thickness are labeled, $d_{SC}$ and $d_H$, respectively, and $\Delta e\Psi_H$ the potential drop in the Helmholtz-layer. $E_{VAC}$ and $E^O_{fb}$ denote the vacuum level and the flatband potential without additional charge in the surface state, respectively and ξ is its shift induced by charging from dipole effects or adsorbates. (a) For energetic alignment of the charge neutrality level (CNL) with the redox potential of the electrolyte, no electrostatic double-layer potential drop (see $\Delta e\Psi_H$) develops. (b) After applying a potential, the space charge layer mainly remains unaffected and the additional exchange charge (see $Q'_{SS}$) leads to a potential drop $\Delta e\Psi_H$ across the double layer. (c) A similar effect can also happen when the surface states are charged during illumination for slow charge-transfer to the electrolyte (dynamic Fermi-level pinning). Quasi-Fermi levels for electrons (holes) are shown in blue (red).*

Due to the involved reactivity, to be expected for probably all multi-electron transfer reactions at low-bandgap semiconductors (not allowing the formation of radicals as intermediates), additional surface modifications are usually operative, which ranges from addition of co-catalysts up to application of passivation layers[3,17,18,2,19]. As a consequence, the junction becomes even more complex as these additional layers must again be adjusted in their energetic alignment to avoid interfacial losses under operation conditions. Again depending on the possibly involved surface reactions with charge carriers provided by the semiconductor during contact formation, the sc/el interface will be further modified under operational conditions. Such intermediate layers, as sketched in Fig. 3, can be formed intentionally or unintentionally in the electrolyte or by separate surface treatment steps, such as thin film synthesis, as also discussed by Nielander et al in their taxonomy paper on photoelectrosynthetic devices[1]. The corrosion-induced formation in the electrolyte, be it intentional or not, actually bears resemblance to the solid-electrolyte interphase formation in batteries[20]. There, it has been shown that this process is highly relevant for application case and consequently, the community dedicates significant efforts to understand and control the formation process. We do not want to claim that a rigorous application of all these considerations will always help solving the efficiency issues of all fuel-forming photoelectrode systems. However, the distribution of electronic states and their involvement in charge carrier trapping, transfer, and interfacial reactions across the complex interfacial sequence of phases are the dominating influence parameters for systems without severe bulk limitations. Therefore, they must be adjusted accordingly for a competitive photoelectrosynthetic electrode arrangement. One example for such an arrangement is show in Fig. 4, where charge-carrier recombination is reduced by a sequence of two *window layers*, also reducing (electro)chemical corrosion.

## III. Combining experimental and computational methods towards a more in-depth understanding

For the design of the photoelectrosynthetic interface with its above-mentioned boundary conditions and optimization requirements, insights on an atomistic level on the interrelation between real-space composition, structure, and electronic structure and their dynamic modifications in the domains of time, applied potential, and space have to be gained. In principle, this covers time-scales spanning about 20 orders of magnitude, from the thermalisation of excited charge-carriers ($10^{-14}$ s)[21] to slow electrochemical corrosion processes. In the following, however, we will focus on the electrochemical timescales of about $10^{-3}$ to $10^2$ s, which are related to the structural reordering of the interface during a potential sweep or upon the onset of illumination. The most relevant properties here for efficient photosynthetic devices are those that are related to the quasi-Fermi level splitting and charge-transfer, i.e. chemical composition, the formation of surface states, band alignments, adsorbed ions, and structure of the EDL.

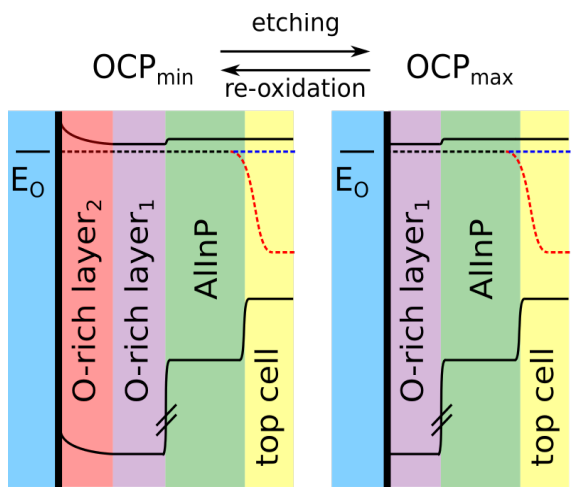

*Figure3: Dynamically changing surface composition of an AlInP-terminated tandem under illumination at open-circuit potential (OCP)[3]. A slow etching process in the aqueous electrolyte (potential $E_0$) leads to build-up and removal of a 0.5 nm thick oxide layer, associated with maxima and minima of the measured OCP. Here, a variation of the CNL is assumed to be the reason for the change in OCP.*

Experimentally, the presence of the liquid electrolyte imposes the challenge to transport information on the (electronic) structure through a solid or liquid with sufficient spatial (ideally Å) and temporal (ms) resolution. Here, the recent developments in near-ambient pressure X-ray photoelectron spectroscopy (NAP-XPS) have been a prominent success story, providing insights on the electronic structure under nearly (very thin electrolyte layer, small currents) realistic conditions. These advances have been covered by a multitude of recent reviews[22,23,24,25]. A notable example illustrating the dynamical change of the interface under operating conditions is the discovery of a reversibly formed, light-induced electronic passivation layer between $BiVO_4$ and a potassium phosphate electrolyte[26]. This also demonstrates the challenge the community faces in designing efficient photoelectrosynthetic devices, where electronic surface passivation is more challenging than in photovoltaic devices due to the dynamical electrochemical response of the interface. Another example is sketched in Fig. 3, showing the band diagram of the topmost layers of a tandem cell in contact with an aqueous electrolyte during surface functionalisation, prior to catalyst deposition. A periodically changing open-circuit potential (OCP) under illumination was observed with a period of about 20 s, associated with the build-up and removal of a thin oxide layer[3]. The maximum of the OCP was associated with a minimum oxide layer thickness and vice versa. The origin of the OCP change could be a different CNL for the two layers as in Fig. 2b, or a different recombination rate (Fig. 2c). The exact nature of the underlying process and the associated energy diagrams is still unclear, also due to the narrow time and process parameter window associated with the oscillation. Here, NAP-XPS could be instrumental in a clarification.

Yet a single method is typically not sufficient to unravel all relevant properties of the interface, which is why complementary methods are employed. These can be of experimental nature, like electrochemical scanning tunneling microscopy or surface X-ray diffraction to probe the spatial structure[27,28]. Increasingly, this is can also be achieved by combining experimental and computational approaches as density functional theory (DFT) based calculations are becoming more apt in describing electrochemical systems. This is due to advances in the theory for the description of interfaces under applied potentials, improved (or now affordable) functionals, and increased feasible system sizes due to advancements in the available computational power[29,30,31,32]. Still, spatial dimensions that can be covered under 3D-periodic boundary conditions typically lie in the order of a few nm$^3$ or hundreds of atoms. In the time domain, the dynamics of the electrolyte is modelled by DFT molecular dynamics (DFTMD), but with a typical time-step in the order of femtoseconds, accessible trajectories lie in the order of picosecons[30]. The computationally less expensive, surface-science approach of only a few monolayers of electrolyte on top of the electrode will only deliver limited insights. A further challenge is related to the fact that DFT is a ground-state method, which does not directly provide the quantities measured by experiment. For systems without strong many-body effects, experimental XPS data can be interpreted in context with DFT(MD) with limited additional computational effort [33]. Computational optical spectroscopy, on the other hand, increases the costs by at least one order of magnitude, even in the case of a linear spectroscopy, like electrochemical reflection anisotropy spectroscopy (RAS)[34,35,36]. Combined with the necessity to perform such an excited state calculation on top of MD snapshots for averaging, this is an expensive, but rewarding undertaking, as it can enable the quantitative interpretation of experimental spectra[36].

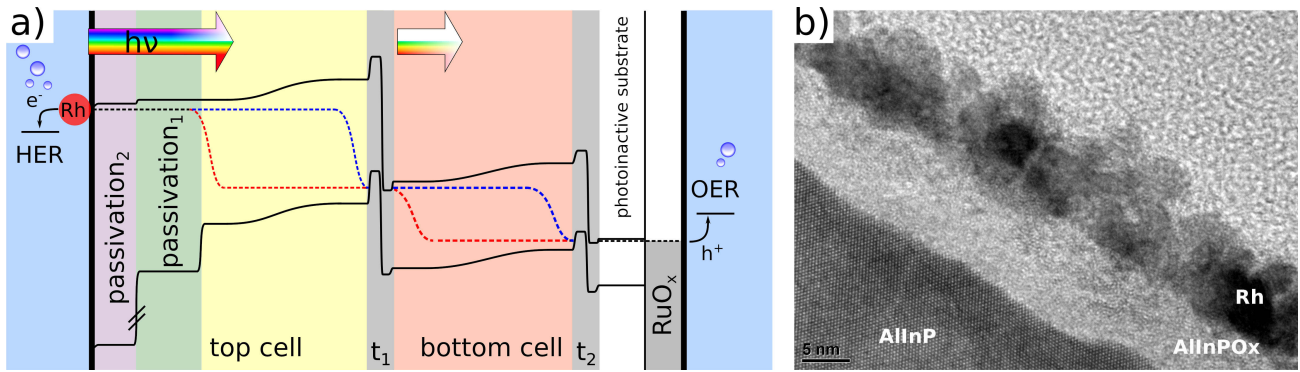

*Figure 4: Surface passivation stack of a III-V dual junction cell for solar water splitting[3] after full functionalisation. a) Band diagram with quasi-Fermi levels for electrons (blue) and holes (red), tunnel junctions (t1,2) and passivation layers. Passivation layer1 serves as purely electronic passivation, while layer2 also passivates the surface chemically and couples to the catalyst nanoparticles. Due to their transparency, these layers are often referred to as window layers. b) Transmission electron microscopy of the topmost layers. The task to understand and control all interfacial properties is in principle more facile for such an epitaxial, buried-junction device, where space-charge layers are less extended and/or further away from the interface, yet also here, it has not been fully achieved, yet.*

This leads us to a final, but very important point on the interpretation of experimental data from advanced methods, especially when performed in context with computational data, where the latter was necessarily generated from highly ordered surfaces: surface quality and reproducibility. Already small amounts of contaminants or a slightly different step edge density on a single-crystal surface can change the interaction with water and subsequent formation of extrinsic surface states qualitatively and quantitatively[37], which can then compromise the relation to computational data. Different surface oxide species can then lead to shallow or more detrimental midgap states[38],[39]. However, even basic information on sample-offcut of single crystals, which determines step edge densities, is often not disclosed in experimental literature. Furthermore, the availability of high-quality single-crystals is limited for many emerging materials. This renders translating insights from model-experiments to real devices, that is already challenging for epitaxial structures[3] as shown in Fig. 4, even more ambitious. In terms of reproducibility, the application of quantitative, highly interface-sensitive probes of the surface structure/quality, such as RAS, prior to or during elaborate NAPXPS experiments would be ideal. This, together with the re-use of data enabled by an increasing number of FAIR data publications, could benefit the efficient interaction between experimental and computational methods, which is necessary to accelerate developments in the field leading to exploitable insights.

## IV. Conclusions

In summary, we feel safe to argue that for semiconductor absorber layers as the light-harvesting component of the device structure, which is in contact with the electrolyte, interfacial effects are a crucial part of the solar fuel generator. For controlling these interfaces on the way towards high-performance devices, they need to be understood in their electronic properties before their use, in operation, and post mortem. Depending on the involved interfacial reactions, the starting configuration may not at all be valid for the operation conditions due to structural dynamics and may induce considerable changes during use and possible degradation. These effects depend on type, size, synthesis, and processing sequence of the electrodes. Currently, neither experimental or computational methods alone are, at least for most systems, able to deliver a full atomistic view of these dynamical, complex interfaces. Advances in a given technique will certainly increase the hereby accessible level of insight, but most techniques are limited to one aspect, e.g. to measure electronic states, but not the real-space interfacial structure. Consequently, an apt combination of theoretical, computational, and experimental approaches is needed. This ranges from the characterisation of well-defined and properly chosen model systems – preferentially using single crystal surfaces – to performance-optimized photoelectrochemical interfaces and finally including in-situ and in-operando techniques with high time resolution. We are convinced that additional advanced spectroscopic and theoretical approaches will develop in the near future applying optical, electronic, and other sample probes. These will, as we believe, ultimately enable the community to gain the needed insight in the chemistry, structure and electronic structure of semiconductor/electrolyte interfaces. With this knowledge, it will be possible to define the design considerations and realize the interface engineering needs for achieving high-performance photoelectrosynthetic devices.

## Declaration of Interests

None.

## Acknowledgements

The authors would like to thank the many colleagues, with whom they have led inspiring discussions on the topic over the last years and U. Bloeck (HZB) for the TEM measurement. WJ acknowledges funding from the German Research Foundation (DFG) project numbers 424873219 and 424924805, MM from DFG project number 434023472 and the German Bundesministerium für Bildung and Forschung (BMBF), project "H2Demo" (No. 03SF0619K).